\documentclass[preprint]{aastex}
\usepackage{psfig}
\usepackage{epsfig}
\usepackage{natbib}
\bibpunct{(}{)}{;}{a}{}{,} 

\shorttitle{Resonance in accreting pulsar}
\shortauthors{Klu\'zniak et al.}

\begin{document}
\title {Non-linear resonance in the accretion disk of a millisecond pulsar}

\author{Wlodek Klu\'zniak\altaffilmark{1,2,3},
 Marek A. Abramowicz\altaffilmark{4},
 Shoji Kato\altaffilmark{5},
 William H. Lee\altaffilmark{6},
 Nikolaos Stergioulas\altaffilmark{7}}


\affil{$^1$Institute of Astronomy, Zielona G\'ora University,
 ul. Lubuska 2, 65-265 Zielona G\'ora, Poland\\ $^2$Centre d' Etude
 Spatiale des Rayonnements, CNRS/UPS, 9 Avenue du Colonel Roche, 31028
 Toulouse Cedex 04, France\\ $^3$Copernicus Astronomical Center,
 ul. Bartycka 18, 00-716 Warszawa, Poland\\ $^4$Astrophysics
 Department, Chalmers University, S-41296 Goteborg, Sweden\\
 $^5$Department of Informatics, Nara Sangyo University, Ikoma-gun,
 Nara, 636-8503, Japan\\ $^6$Instituto de Astronomia, UNAM,
 Apdo. Postal 70-264 Cd, Universitaria, DF 04510, Mexico\\ $^7$Physics
 Department, Aristoteleion University of Thessaloniki, 541.24
 Thessaloniki, Macedonia, Greece}


\begin{abstract} Twin quasi-periodic millisecond modulations of the
X-ray flux (kHz QPOs) have recently been reported from an accreting
2.5 ms X-ray pulsar.  We identify modes of disk oscillations whose
frequencies are in agreement with the observed ones
 when the rotating neutron
star is modeled with realistic equations of state.  The frequency
difference of the twin QPOs, equal to about one
half of the neutron-star spin rate, clearly indicates that resonant
oscillations of the accretion disk have been observed. Similar
non-linear resonances may also be spontaneously excited in the
accretion disk. The two QPO frequencies in the pulsar system are close
to a 5:7 ratio and this suggests a link with the QPOs in black hole
systems, where frequency ratios of 2:3 and 3:5 have been reported.
\end{abstract}


\section{Introduction}

A millisecond pulsar that accretes matter from a binary stellar
companion provides a unique probe of the physics of accretion flow a
few Schwarzschild radii away from the neutron star. In contrast with
the case of the slowly rotating steady X-ray pulsars, the magnetic
field of the (X-ray transient) millisecond pulsar is too weak to
prevent the accreting gas from orbiting the neutron star very close to
its surface. Hence, in such a pulsar oscillations in the inner parts
of the disk may be observed at their characteristic frequency of
several hundred Hertz.  The pulsar is expected to disturb the
accretion disk at its spin frequency. That it does, is demonstrated by
the discovery in the 2.5 ms accreting pulsar of a frequency difference
between the  two QPOs equal to one-half the pulsar spin rate.

We point out that the two QPO frequencies themselves may be  commensurable
with the spin frequency and with each other. The two simultaneously
observed QPO frequencies are close to 500 Hz and 700 Hz, i.e., they may be
in a 5:7 ratio. Twin QPOs have been proposed to be
 a manifestation of  non-linear resonance in relativistic
accretion disks \citep[]{66}, and the resulting suggestion that
twin QPO frequencies should be in the ratio of small integers
has subsequently been confirmed for a number of black hole transients
\citep[]{6,7,8,22} and for the bright steady source Sco X-1 \citep[]{23}.

The two QPOs discovered in the 2.5 ms accreting pulsar \citep[]{1} are
similar to QPOs previously detected in many unpulsed X-ray sources, in
which an accretion disk is thought to feed matter and angular momentum
to the neutron star or black hole \citep[]{2,3}. Such QPOs in black
hole systems were intrerpreted as oscillations of the accretion disk
\citep[]{4,5}, but a direct proof of this was lacking, and for neutron
star systems other models had been proposed \citep[]{2}.
SAX~J1808.4-3658 is the first kHz QPO source in which the rotational period of
the neutron star has been unambiguously determined, and the first in
which the frequency difference of the twin QPOs is known to be a
subharmonic of the stellar rotation rate. For the first time, 
we have direct evidence that kHz QPOs are caused by disk oscillations.

\section{Non-linear resonance and epicyclic frequencies}
The presence of a sub-harmonic frequency is a clear signature of
non-linear resonance \citep[]{9}. Theodor von Karman observed in 1940 that
high-frequency vibrations of an airplane engine excite lower-frequency
resonances in the airframe \citep[]{10,9}. A tragic example is known when a
subharmonic resonance was excited in an airplane wing, which in turn
excited a subharmonic resonance in the rudder, at 1/2 the wing
eigenfrequency \citep[]{11,9}. In the present case, the 2.5 ms pulsar SAX
J1808.4-3658 plays the role of the engine, while the accretion disk
(whose response is observed as the quasi-periodic oscillations) is the
airframe with its own set of known eigenfrequencies. The fact that the
QPO frequency difference coincides with one half the known pulsar spin
frequency,  
$$(694\pm4)\,{\rm Hz}-(499\pm4)\,{\rm Hz}=
(195\pm6)\,{\rm Hz}\approx{1\over2}(401\, {\rm Hz}),$$
 follows directly from
the nature of the system. The structure of the accretion disk is
determined by the non-linear equations of hydrodynamics, and the
effective gravitational potential (in which the accreting gas moves)
is not harmonic, so a non-linear response is expected. Below, we
discuss in detail how this comes about for an accretion disk in the
space-time of a rotating neutron-star.

The motion of a test particle in nearly circular orbits close to the
equatorial plane can be decomposed into three components, circular
planar motion at the orbital frequency $\Omega\equiv2\pi\nu_{\rm orb}$,
harmonic radial motion at the radial epicyclic frequency 
$\kappa\equiv2\pi\nu_{\rm r}$, and harmonic vertical motion at the
meridional epicyclic frequency $\zeta\equiv2\pi\nu_{\rm vert}$.
 In Newtonian theory of spherically
symmetric gravitating bodies the three frequencies coincide, but in
Einstein's gravity $\kappa<\Omega$.
 (The relativistic precession of the perihelion
of Mercury occurs at the rate $\Omega -\kappa$.) 
The same three frequencies are
important in a discussion of fluid motion about a gravitating body in
general relativity \citep[]{12}. The accretion disk is a body of hot gas that
is supported against infall primarily by rotation and is nearly in
hydrostatic equilibrium \citep[]{3,12}. Theories of accretion 
\citep[]{13,14} admit
solutions in which the disk thickness is much smaller than its radial
extent, but also solutions in which the disk has the geometry of a
torus. Like other extended bodies in equilibrium, the disk is capable
of motion in a variety of modes. In the linear regime, small
oscillations of geometrically thin accretion disks, as well as waves
in this body of fluid, have been extensively studied in the Kerr
metric \citep[]{4,5}. The radial vibrations of two-dimensional models of
geometrically thick disks (accretion torii) have been investigated
numerically in the Schwarzschild metric \citep[]{15}. Qualitatively, these
black-hole metrics are similar to the metrics of neutron stars.

We have studied numerically the response to a transient external
perturbation of an ideal-gas torus that is initially in equilibrium
rotation about a (Schwarzschild) black hole or neutron star. We find
that the torus performs harmonic oscillations both in the radial and
vertical directions, even if the impulse imparted to it at the
beginning of the computation is purely radial. Further, the vertical
oscillations have variable amplitude. Both these effects speak of a
non-linearity in the system.The vertical and radial displacements of
the center of the torus (defined as the point of maximum pressure in
a meridional cross-section) are shown as a function of time in
Fig. 1. The radial displacement has been rescaled to fit the graph---in
the simulation described here, the amplitude of radial motion is actually
 larger than the amplitude of vertical motions.

\begin{figure}[!ht] \plotone{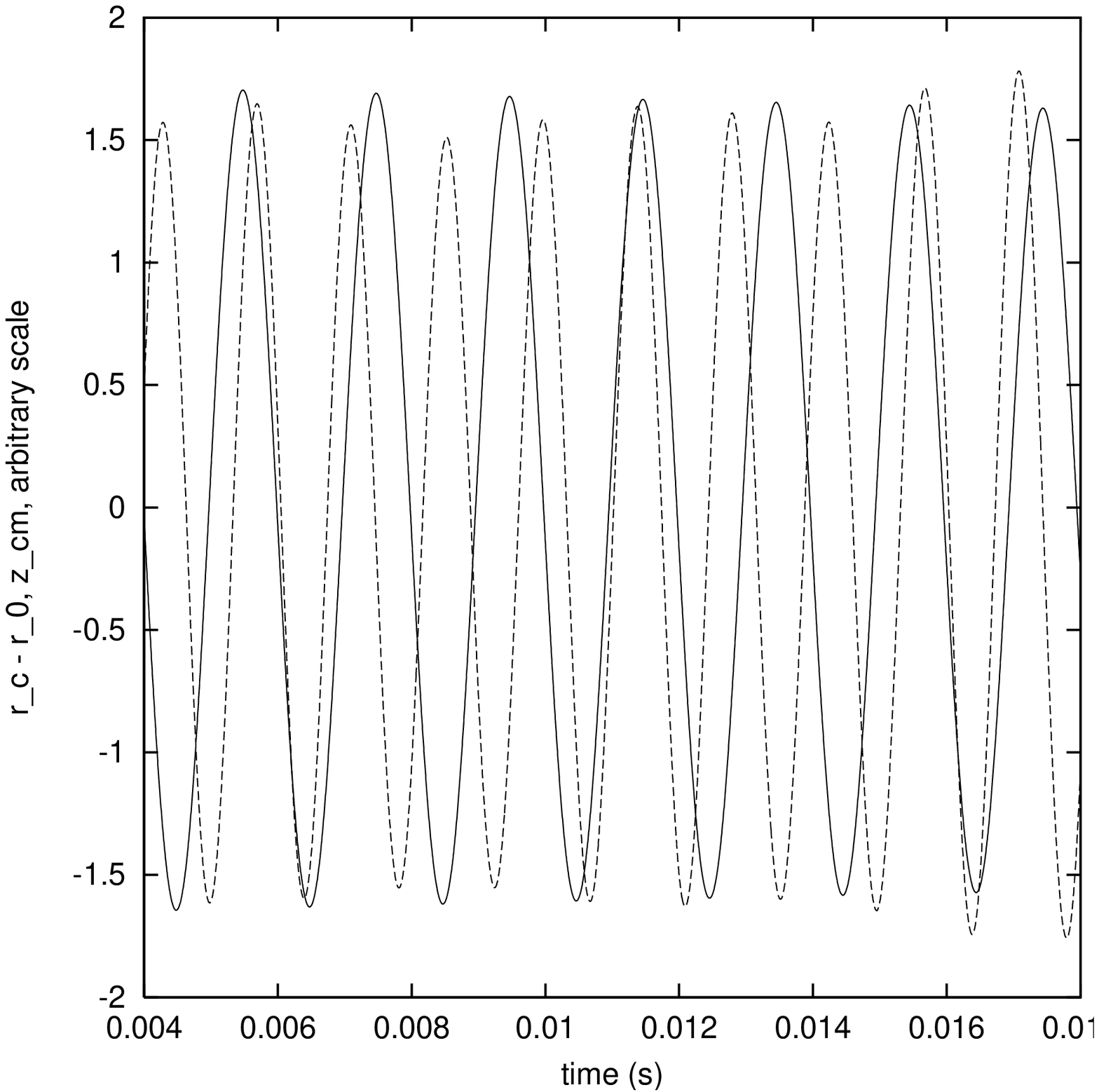}
\caption[]{The vertical and radial motions of an accretion torus in
response to an impulsive radial perturbation at time $t=0$. The vertical
displacement of the torus (dashed line) about the equatorial plane and
the radial displacement (continuous line) about the equilibrium
radius, $r_0$, of the circle of maximum pressure are plotted as a function
of time. In this pseudo-Schwarzschild simulation, $r_0=12.25M (G/c^2)$. It is
clear that the torus performs two harmonic motions. We find that their
frequencies coincide with the meridional and radial epicyclic
frequencies, $\nu_{\rm vert}$ and $\nu_r$,
 here 700 Hz and 500 Hz, respectively. The ratio of
the frequencies is a function of $r_0$, their value a function of the mass
$M$ of the neutron star. See the text for details.}
\label{fig1}
\end{figure}

For the
numerical simulation (Fig. 1), we use a Newtonian 2-d smooth particle
hydrodynamics \citep[SPH]{16} code \citep[]{17} simulating ideal gas with
adiabatic index 4/3 in a pseudo-potential  \citep[]{18}
$\psi_{KL}(r)=[1-\exp(r_{\rm ms}/r)]GM/r_{\rm ms}$, which reproduces the
Schwarzschild ratios of the orbital and epicyclic frequencies,
$\zeta/\Omega=1$, $(\kappa/\zeta)^2=1-r_{\rm ms}/r$.
 Here, $r_{\rm ms}=6M (G/c^2)$ is the radius of the marginally stable orbit.
 A torus in equilibrium, with $r_0=12.25M (G/c^2)$, 
was perturbed at time $t=0$ by imparting to it a
radial velocity field with magnitude proportional to $\sqrt{r_{\rm ms}/r}$.
 The frequencies
of the resulting motions are found to be inversely proportional to the
central mass $M$, whose value was adjusted to match the observed QPO
frequencies.

The two frequencies observed in this numerical simulation coincide,
within errors, with the known epicyclic frequencies, 
$\kappa$ and $\zeta$, at the
equilibrium radial position, $r_0$, of the locus of maximum pressure, $r_c$.
 We conclude that a slender torus responds to a radial
external perturbation with oscillations occuring at
frequencies  equal to the two
epicyclic frequencies, $\kappa (r_0)$ and $\zeta(r_0)$,
at that radius where the pressure of the torus
in equilibrium is highest. An important implication is that these
frequencies can vary if the torus varies with time as it
accumulates less or more of the matter flowing through it.

For the radial perturbation applied in the simulation, the amplitude
of vertical motion is smaller than that of the radial
motion. Presumably the ratio of the amplitudes depends on the
perturbation applied. Quantifying this requires further study. A
tilted magnetic dipole may perturb the disk quite strongly in the
vertical direction. In the remainder of this section we assume
that the perturbation is such that only one of the two motions is typically
excited with a large amplitude, and that the other would not be as easily
discernible in an X-ray observation, unless its initially much smaller
amplitude were amplified by a resonance.

A pulsar is a rotating neutron star with a strong magnetic dipole not
aligned with the rotation axis. As a result, in the 2.5 ms accreting
pulsar \citep[]{19} the accretion disk suffers a periodic disturbance at the
spin frequency $\nu_1 =401\,$Hz. Observations indicate the presence of a QPO
varying in frequency between 280 and 750 Hz \citep[]{1}. 
We interpret
this as one of the two epicyclic frequencies, $\nu_0$,
and expect that a second QPO will be
present at the second epicyclic frequency, when
 a resonance occurs at frequency $\nu_{\rm res}$ between the
corresponding disk oscillation and the disturbing forces. 

Twin QPOs have been observed in this source
on only one occasion, and it is not clear to us
whether the single variable frequency  $\nu_0$
present in the remaining observations
corresponds to the upper or the lower QPO, i.e., to $\nu_{\rm vert}$ or to
$\nu_r$ (but see below). 
 A simple and general
reason why the difference is equal to one half the spin frequency,
 $\nu_{\rm res} -\nu_0=\nu_1/2$, and not the spin frequency,
 $\nu_1$, can be given
if $\nu_0=\nu_r$. Then
 $\nu_{\rm res}=\nu_{\rm vert}$ at a particular radius, and the resonance
would occur in an oscillator without quadratic anharmonicity terms.
The restoring force in
the vertical motion of the torus is obtained by taking the  $z$ derivative
of the effective potential expanded about the equatorial
plane. Because the potential is symmetric under reflection in this
plane, this expansion has only even powers of the cylindrical
co-ordinate $z$, $V(z)=V_0+(\zeta z)^2/2 +\beta z^4/4 +$...
The vertical motion of the torus can then be
described as that of an oscillator with (angular) eigen-frequency $\zeta$,
and a cubic anharmonicity in leading order, $\beta z^3$. 

Consider resonance in an accretion disk which can be described as a
 damped oscillator with a cubic anharmonicity.
In view of the very
strong disturbance by the pulsar, the resonance may be driven by a
combination of the pulsar spin frequency $\nu_1$ and another, much weaker,
disturbance at frequency $\nu_2$, which we take to be a harmonic (i.e., an
integer multiple) of the QPO frequency $\nu_0$. 
 In the present case the frequencies of the two harmonic driving forces 
satisfy $\nu_2>\nu_1$. 
The oscillator may be in resonance with both of these forces at
the same time \citep[simultaneous resonance]{9} in one of three cases:

a) $\nu_{\rm res}\approx 2\nu_1\pm \nu_2$, or 

b) $\nu_{\rm res}\approx 2\nu_2\pm \nu_1$, or 

c) $\nu_{\rm res}\approx(\nu_2\pm \nu_1)/2$.

\noindent  The single occasion
when two QPOs were detected in the 2.5 ms pulsar corresponds to case
c), with $\nu_2=2\nu_0$, i.e., $\nu_{\rm res}\approx\nu_0\pm \nu_1/2$.
Since for neutron stars in general relativity $\nu_{\rm vert}>\nu_r$,
the upper sign corresponds to the case
$\nu_{\rm res}=\nu_{\rm vert}$, $\nu_0=\nu_r$, the lower sign to the case
$\nu_{\rm res}=\nu_r$, $\nu_0=\nu_{\rm vert}$.
 Indeed,  the two observed QPO frequencies, are
in this relation to the spin frequency of 401 Hz, 
$\nu_{\rm upper}-\nu_{\rm lower}=694\,{\rm Hz}-499\,{\rm Hz}
\approx(401\,{\rm Hz})/2=\nu_1/2$.

 We note that the
two QPOs cannot be separated by the spin frequency in this
scheme. A separation between the two QPO 
frequencies  of twice the spin frequency,
($\nu_{\rm res} - \nu_0=802\,$ Hz, case a with $\nu_0=\nu_2$)
is in principle also possible in this scheme if
resonance occurs at the very inner edge of the disk, at frequencies of
about 650 Hz and 1450 Hz, but it does not seem likely that the bulk of
the torus could reside so close to the neutron star.

\begin{figure}[!ht] \plotone{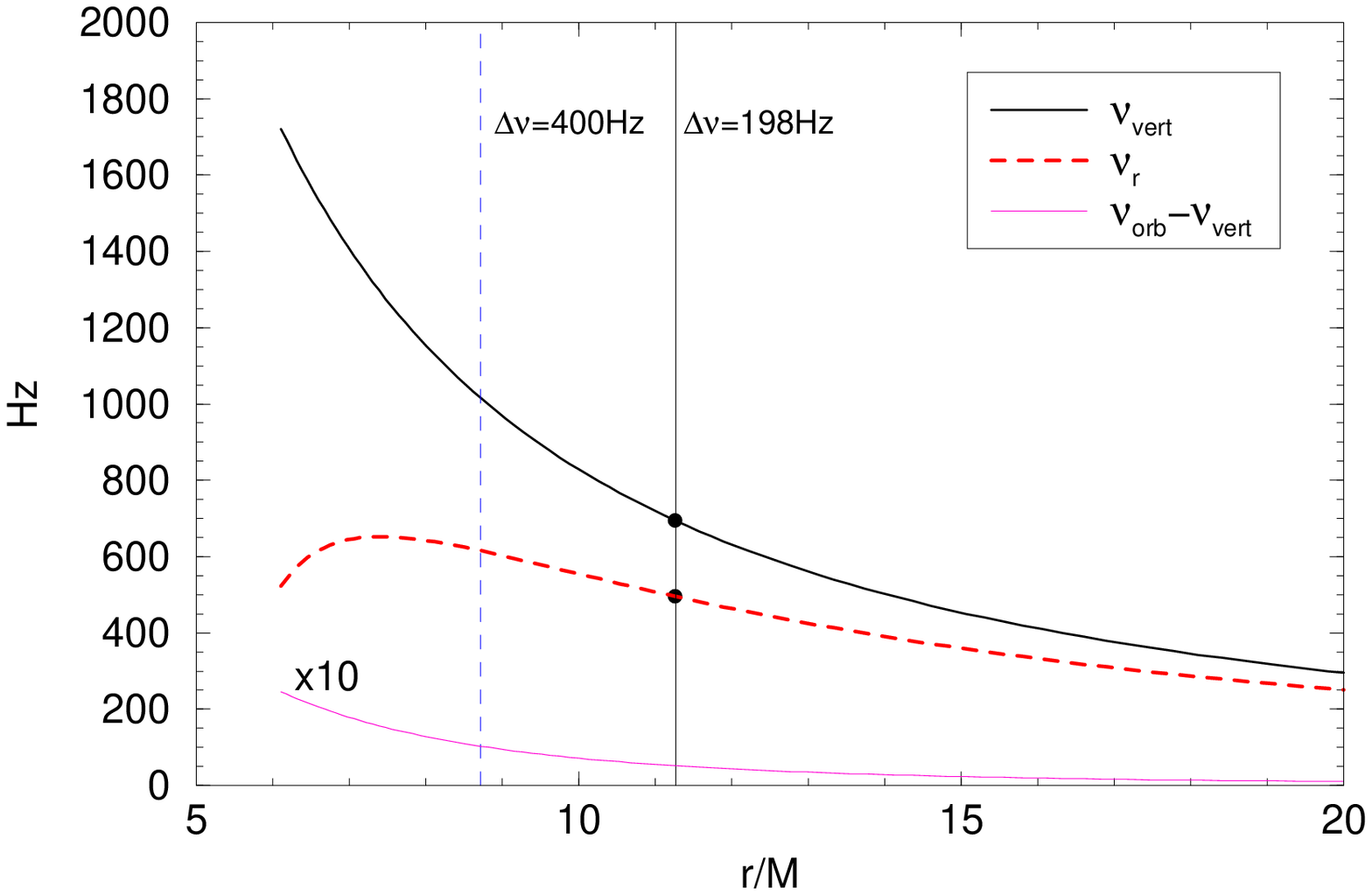}
\caption[]{The radial and vertical epicyclic frequencies 
for a neutron star
rotating at a rate of 401 Hz. The two thick dots indicate where the
two epicyclic frequencies are separated by 1/2 the spin frequency
(within grid errors). A torus whose circle of maximum pressure has
radius $r_0=11.27M$
 (vertical line) would oscillate at frequencies 496 Hz and 694
Hz. The vertical dashed line indicates the radius 
($r=8.72M$) where the
epicyclic frequencies are separated by the spin frequency. Also shown
is ten times the Lense-Thirring frequency. The neutron star was
modeled with equation of state FPS and has 1.22 solar mass. All
frequencies are plotted as a function of the radius, given in units of
$M(G/c^2)=1.22\times1.5\,$km. 
\label{fig2}}
\end{figure}

We have numerically computed realistic models of rotating neutron
stars and their exterior metrics.
 Fig. 2 presents a typical result consistent with observations
of the two QPOs in the 2.5 ms pulsar. The vertical and radial
epicyclic frequencies for a 1.22 solar mass neutron star, modeled with
equation of state FPS, and rotating at 401 Hz are illustrated as a
function of the radius. The case c) resonance discussed above occurs
when pressure in the torus has a maximum at radius 
$r_0=11.27M(G/c^2)=20.6\,$km. The
stellar mass and resonance radius are only indicative of the true
values for this pulsar, as the frequencies for a flattened torus of
large radial extent may differ from $\kappa(r_0)$ and $\zeta(r_0)$
 by several per cent \citep[]{15}.

Wijnands et al. (2003) give phenomenological
arguments in favor of the frequency being the upper QPO.
In this case, in addition to the resonance described below,
 a resonance at the combination frequency
$\nu_r =\nu_{\rm vert}-\nu_1$ could also
occur in principle
 if the radial oscillator with its {\it quadratic} anharmonicity were
perturturbed simultaneously at the spin frequency and the frequency of
the vertical oscillation. This would happen at $r_0=16\,$km for the same
model of the neutron star, when the upper QPO frequency has the 
(as yet unobserved value) 1015 Hz  and the lower frequency is 615 Hz
(dashed vertical line in Fig. 2).

To compute the frequencies
in Figs.~2 and 3, we constructed numerical models of rotating neutron
stars and their exterior metrics using the relativistic code of
Stergioulas and Friedman \citep[]{20}, which solves Einstein's field equations
for arbitrarily large rotation rate in an integral form. Details of
the numerical method and extensive accuracy tests can be found in
 \citep[]{21}. The results presented here are not very sensitive to the
choice of equation of state (EOS) of neutron-star matter.

\section{Parametric resonance}
So far we have been discussing a resonance directly caused by an
external disturbance. But when the frequencies $\kappa(r)$
 and $\zeta(r)$ happen to be
in particular ratios, large-amplitude motions of the torus may be
excited spontaneously. This mechanism, and specifically parametric
resonance between the two modes of oscillation, has been suggested 
\citep[]{8}
as the origin of the high-frequency QPOs in black hole systems, where
the frequency ratios 5:3 and 3:2 have been noted \citep[]{6,7,8,22}, as well
as in the candidate neutron-star system Sco X-1 \citep[]{23,24}. The
presence of subharmonics in some of the black hole systems has also
been discussed \citep[]{7,25}. On the other hand, in several X-ray bursters
the frequency difference between the twin QPOs has been reported to be
close to the suspected spin frequency (or one half of it) of the
neutron star \citep[]{2}. It is interesting to note that in addition to being
separated by one-half the pulsar spin frequency, the two QPO
frequencies  in SAX J1808.4-3658 are approximately in a 7:5=1.4
ratio. The system may thus be the Ômissing linkÕ between the QPOs in
black hole systems and X-ray bursters. It is known that if the
frequency difference of two modes is accidentally close to that of an
external disturbance, a resonance involving the three frequencies may
occur \citep[]{26}. In neutron star systems this could favor the appearance of
strong QPOs with a frequency difference selected by the perturbing
neutron-star rotation, while in black hole disks only the spontaneous
resonance would occur.

\begin{figure}[!ht] \plotone{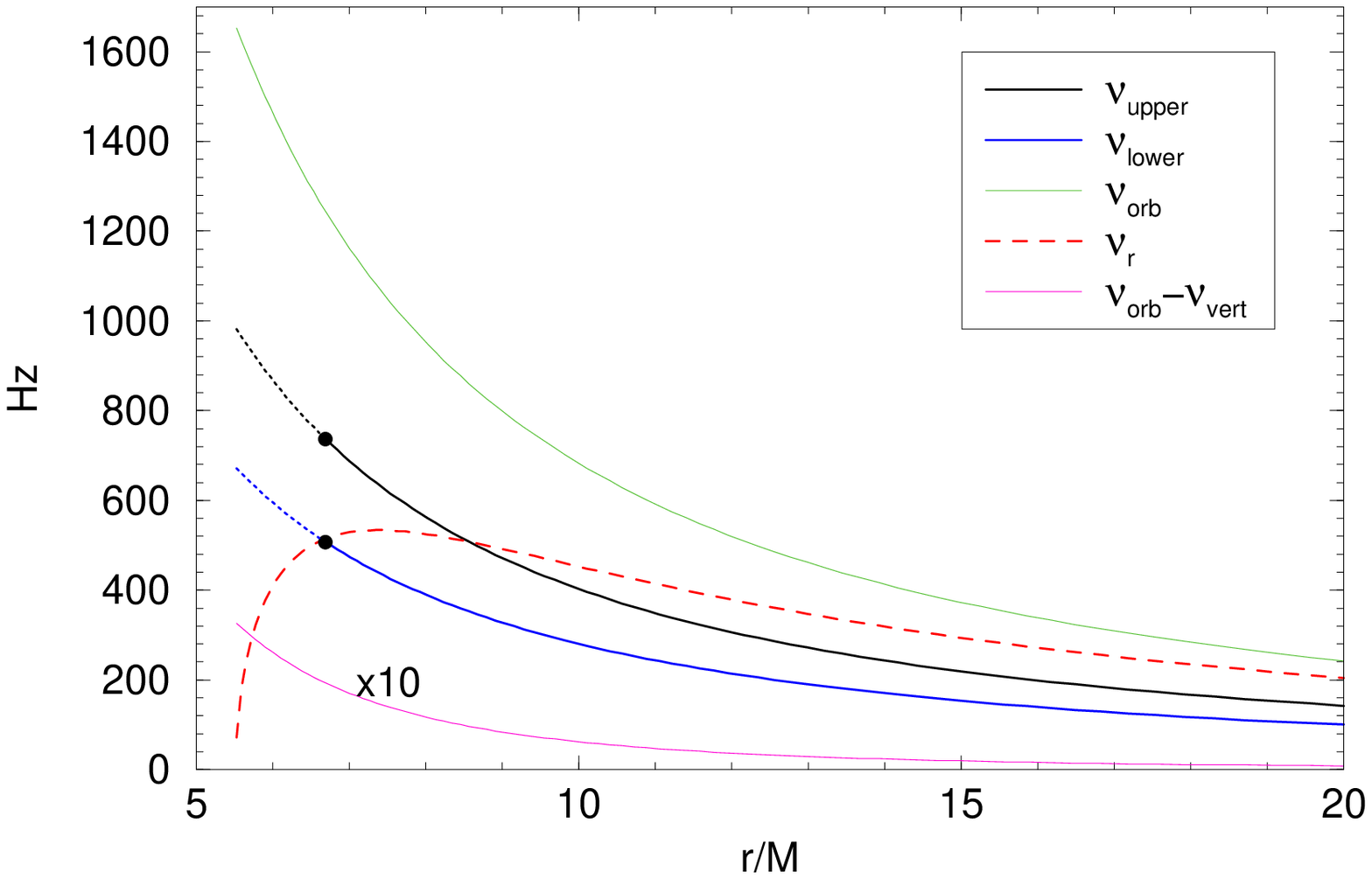}
\caption[]{The frequencies (thick curves) of two
{\it g}-modes of oscillation excited by parametric resonance in a thin
accretion disk. The two thick dots at  $r=6.69M$ mark the boundary of the
instability region---the resonant interaction with a rotating warp
occurs to the right of the dots, where the lower frequency is less
than the radial epicyclic frequency (dashed curve). The largest
modulation of X-rays is expected close to the indicated boundary of
the resonance region, i.e., at about 508 Hz and 737 Hz for the model
computed (1.49 solar mass neutron star rotating at 406 Hz, EOS
AU). For a star of lower mass,  $\nu_{\rm lower}$ and $\nu_{\rm upper}$
 would have the values 500 and 700
Hz at a larger value of $r/M$.The computed frequencies of the two modes are
uncertain by several percent because of the unknown rotation rate of
the warp, which coincides at a poorly constrained radius with the
Lense-Thirring frequency (also shown, in ten-fold magnification).
\label{fig3}}
\end{figure}

Parametric resonance between two oscillations is possible in accretion
disks because the coupling of modes is non-linear in
hydrodynamics. For thin disks, this can lead to excitation of two
modes whose frequency ratio is closer to 1.4 than to 1.5. In the
Schwarschild metric, the damping or excitation of {\it g}-modes with
azimuthal dependence $\exp(im\phi)$ has been considered by Kato \citep[]{27}, 
who found
that an $m=0$ mode and an $m=1$ mode will undergo resonant amplification
when interacting with a non-rotating one-armed ($m=1$) stationary warp,
assumed to be present in the accretion disk. We assume that the same
modes
 are unstable also in a gravitational field that is not
spherically symmetric, and compute their frequencies for the metric of
rotating neutron stars. The sum of the two frequencies turns out to be
$\nu_{\rm orb}$,
 and if they are identified with the two observed QPO frequencies
this constrains the mass of the 2.5 ms pulsar to be less than 1.5
solar mass. The two mode frequencies,  
$\nu_{\rm lower}$ and $\nu_{\rm upper}$,
 as well as $\nu_{\rm orb}$, $\nu_r$, and $\nu_{\rm orb}-\nu_r$ are
shown in Fig. 3 for a 1.49 solar mass neutron star, spinning at a
frequency of 406 Hz. The resonance occurs in the region where
$\nu_{\rm lower} <\nu_{r}$, with
the largest modulation of X-ray flux expected when the two frequencies
are evaluated at the edge of the resonance zone, i.e., when they are
$\nu_{\rm lower} \approx\nu_{r}$ and 
$\nu_{\rm upper}\approx\nu_{\rm orb}-\nu_r$.

{The two unstable modes
in Fig. 3 were taken, following \citep[]{27}, to be $m=1$, $n=1$, 
and $m=0$, $n=1$,
{\it g}-modes excited by parametric resonance, which is mediated by a warp
assumed to be present in the thin accretion disk. For a non-rotating
warp their frequencies \citep[]{27} would be 
$2\nu_{\rm orb}-\sqrt{2}\,\nu_{\rm vert}$ and 
$\sqrt{2}\,\nu_{\rm vert}-\nu_{\rm orb}$, respectively, with
their ratio tending to 
$\sqrt2=1.41$ in the limit of non-rotating stars, or in
the limit of large radial distance from the star, $r\rightarrow\infty$.
 The warp can be
identified with a {\it c}-mode of disk oscillations and is expected to
rotate at the Lense-Thirring angular frequency
 $2\pi(\nu_{\rm orb}-\nu_{\rm vert})$, evaluated at a
certain characteristic radius, $r_w$ \citep[]{30}. The two curves plotted in
Fig. 3 are $\nu_{\rm lower}(r)=(\sqrt2-1)\nu_{\rm vert} (r)$
 and $\nu_{\rm upper}(r)=\nu_{\rm orb}+(1-\sqrt2)\nu_{\rm vert}(r)$.
 The actual mode frequencies coincide with these at $r=r_w$,
 and at arbitrary radius  differ from these by 
$[\nu_{\rm orb}(r_w)-\nu_{\rm vert}(r_w)]-
[\nu_{\rm orb}(r)-\nu_{\rm vert}(r)]$, 
i.e., by a few percent. For the present purposes,
we take the quantity  $r_w$ to be an unknown parameter.}

\section{Conclusions}
In conclusion, we note again that it is not yet known whether the
accretion disk resembles a torus or, to the contrary, is geometrically
thin. The 2.5 ms pulsar is the first source in which the frequencies
of the twin QPOs are at once clearly related to the neutron-star spin
frequency and are in a 1.4 ratio. With one observation of twin
frequencies in only one source with a known rotation rate (but an
unknown mass), it is impossible to determine which of the two pairs of
oscillation modes discussed here is responsible for the non-linear
resonance. The ambiguity will be removed if twin QPOs are observed in
one or more of the four other accreting pulsars \citep[]{29} known, where the
spin periods are different.

It is worth noting that the fact that the frequency
difference between the two observed kHz QPOs is equal to one half the
spin frequency of the neutron star is the first clear indication that
the commonly observed millisecond modulations of the X-ray flux 
in low mass X-ray binaries are
caused by oscillatory motions of the accretion disk around the neutron star
(or black hole in other systems).

\begin{acknowledgements}
The authors thank Luciano Rezzolla for pointing out
to them the importance of the radial oscillations of a torus. WK
wishes to thank Prof. Shin Mineshige for generous hospitality at the
Yukawa Institute, and Prof. Jacques Paul for welcoming him at the
Cargese School (IESC). MAA, WK, WHL and NS acknowledge the hospitality
of SISSA. This work has been supported in part by the EU Programme
Improving the Human Research Potential and the Socio-Economic
Knowledge Base (Research Training Network Contract
HPRN-CT-2000-00137), KBN grant 2P03D01424, the Greek GSRT Grant
EPANM.43/2013555, by CONACYT,
and by le Departement Sciences de l'Univers du CNRS.
\end{acknowledgements}


\begin{thebibliography}{}

\bibitem[Abramowicz et al.(2003a)]{23}
	M.A. Abramowicz, T. Bulik, M. Bursa, W. Klu\'zniak,
      Astron. Astrophys. 404, L21-L24 (2003a).  

\bibitem[Abramowicz et al.(2003b)]{24}
	M.A. Abramowicz, V. Karas, W. Klu\'zniak, W.H. Lee, P. Rebusco,
  Publ. Astron. Soc. Jpn. 55, 467-466 (2003b).  

\bibitem[Abramowicz and Klu\'zniak(2001)]{6}
    M.A. Abramowicz, \& W. Klu\'zniak, Astron. Astrophys. 374, L19-L20 (2001).

\bibitem[Jaroszy\'nski et al.(1980)]{14}
	M. Jaroszy\'nski, M.A. Abramowicz, B. Paczy\'nski, Acta
     Astronomica 30, 1-34 (1980).  

\bibitem[Kato(1974)]{26}
	S. Kato, Publ. Astron. Soc. Jpn. 26, 341-353 (1974).  

\bibitem[Kato(1998)]{12}
	S. Kato, S. Mineshige, J. Fukue, Black Hole Accretion
    Disks (Kyoto University Press: Kyoto, 1998).  

\bibitem[Kato(2001)]{5}
	S. Kato, Publ. Astron. Soc. Jpn. 53, 1-24 (2001). 

\bibitem[Kato(2003)]{27}
	S. Kato, Publ. Astron. Soc. Jpn. in press (2003). 
 
\bibitem[Klu\'zniak and Abramowicz(2001)]{66}
	W. Klu\'zniak, \& M. Abramowicz, Phys. Rev. Lett.
    submitted (http://arxiv.org/astro-ph/0105057).

\bibitem[Klu\'zniak and Abramowicz(2002)]{8}
	W. Klu\'zniak, \& M. Abramowicz, Astron. Astrophys.
    submitted (http://arxiv.org/astro-ph/0203314).

\bibitem[Klu\'zniak and Abramowicz(2003)]{25}
	W. Klu\'zniak, M.A. Abramowicz, Proceedings of the 12th Workshop
   on General Relativity and Gravitation. M. Shibata, M., Eriguchi, Y., Eds.,
   $<$http://arxiv.org/abs/astro-ph/0304345$>$ (2003). 

\bibitem[Klu\'zniak and Lee(2002)]{18}
	W. Klu\'zniak, \& W.H. Lee, Mon. Not. R. Astron. Soc. 335, L29-L32
      (2002)

\bibitem[Lee and Ramirez-Ruiz(2002)]{17}
	W.H. Lee, \& E. Ramirez-Ruiz, Astrophys. J. 577, 893-903 (2002).  

\bibitem[Lefschetz(1956)]{11}
	S. Lefschetz, in Modern Mathematics for the Engineer.
      E.F. Beckenback, Ed., pp. 7-30 (McGraw Hill: New York, 1956).  

\bibitem[Lewin et al.(1995)]{3}
	W.H.G. Lewin, J. van Paradijs, E.P.J. van den Heuvel, Eds.,
      X-ray binaries (Cambridge University Press, 1995). 

\bibitem[Markwardt and Swank(2003)]{29}
	C.B. Markwardt, \& J.H. Swank, XTE J1814-338. International
    Astronomical Union Circular 8144 (2003). 

\bibitem[McClintock and Remillard(2003)]{22}
	J.E. McClintock, \& R.A. Remillard,
   http://arxiv.org/abs/astro-ph/0306213 (2003).

\bibitem[Monaghan(1992)]{16}
	J.J. Monaghan, Ann. Rev. Astron. Astrophys. 30, 543-574 (1992). 

\bibitem[Nayfeh and Mook(1979)]{9}
	A.H. Nayfeh, D.T. Mook, Non-linear oscillations (Wiley: New
    York, 1979). 

\bibitem[Remillard et al.(2002)]{7}
        R.A. Remillard, M.P. Muno, J.E. McClintock, J.A. Orosz, Astrophys. J.
  580, 1030-1042 (2002)

\bibitem[Rezzolla et al.(2003)]{15}
	L. Rezzolla, S. Yoshida, O. Zanotti, Mon. Not. R. Astron.
     Soc. submitted (2003).

\bibitem[Shakura and Sunyaev(1973)]{13}
	N.I. Shakura, \& R.A. Sunyaev, Astron. Astrophys. 24, 337-355 (1973).

\bibitem[Silbergleit et al.(2001)]{30}
	A.S. Silbergleit, R.V. Wagoner, M. Ortega-Rodr\'{\i}guez,
    Astrophys. J. 548, 335-347 (2001).

\bibitem[Stergioulas(2003)]{21}
	N. Stergioulas, in Living
Reviews in Relativity,  (2003). 

\bibitem[(1995)]{20}
	N. Stergioulas, J.L. Friedman, Astrophys. J. 444, 306 (1995). 

\bibitem[van der Klis et al.(2000)]{2}
	M. van der Klis, Annu. Rev. Astron. Astrophys. 38, 717-760 (2000).

\bibitem[von Karman(1940)]{10}
	T. von Karman, The engineer grapples with nonlinear problems.
  Bulletin of American Mathematical Society 46, 615-683 (1940).  

\bibitem[Wagoner(1999)]{4}
	R. V. Wagoner, Phys. Rep. 311, 259-269 (1999). 

\bibitem[Wijnands and van der Klis(1998)]{19}
	R. Wijnands, M. van der Klis, Nature 394,
344-346 (1998). 

\bibitem[Wijnands et al.(2003)]{1}
	R. Wijnands, M. van der Klis, J. Homan,
	C.B. Markwardt, E.H. Morgan, Nature in press (2003)
 
\end{thebibliography}
\end{document}